\documentclass[letterpaper, twocolumn]{article} 
\usepackage{times}  
\usepackage{helvet}  
\usepackage{courier}  
\usepackage[hyphens]{url}  
\usepackage{graphicx} 
\usepackage{natbib}  
\usepackage{caption} 
\usepackage{algorithm}
\usepackage{algorithmic}
\usepackage[breaklinks,hidelinks]{hyperref}

\usepackage{xcolor}
\usepackage{soul}
\usepackage{url}

\usepackage{newfloat}
\usepackage{listings}
\DeclareCaptionStyle{ruled}{labelfont=normalfont,labelsep=colon,strut=off} 
\floatstyle{ruled}
\newfloat{listing}{tb}{lst}{}
\floatname{listing}{Listing}
\usepackage{tabularx}
\usepackage{rotating}
\usepackage{multirow}
\usepackage{color}
\usepackage{enumitem}
\usepackage[multiple]{footmisc}
\usepackage{hyperref}

\title{Implications of Current Litigation on the Design of AI Tools for Healthcare Delivery and Related Legal Frameworks}
\author{
    Gennie Mansi~\textsuperscript{\rm 1$\dagger$},
    Mark Riedl~\textsuperscript{\rm 1} \\
    \textsuperscript{\rm 1}
    Georgia Institute of Technology\\
    225 North Avenue NW
    Atlanta, GA\\
    $\dagger$ gennie.mansi@gatech.edu (corresponding author)
}
\date{\vspace{-2.5ex}}

\begin{document}

\maketitle

\begin{abstract}
The medical, legal, and technology communities often center physicians' role in preventing and mitigating harms from medical AI tools.
We conduct an analysis of 31 U.S. legal cases and reported harms to identify patterns around how AI tools impact patient care. Our findings reflect how patients' medical care relies on a complex web of stakeholders---physicians, state health departments, health insurers, care facilities, among others---and many AI tools deployed across their healthcare delivery negatively impact their care. 
In response, patients have had no option but to seek legal recourse for harms.
We shift the frame from physician-centered to patient-centered accountability approaches by describing how lawyers and technologists need to recognize and address where AI harms happen. We present paths for preventing or countering harm
(1) by changing liability structures to reflect the role of many stakeholders in shaping how AI tools impact patient care; and 
(2) by designing tools that can help advocates, such as patients' lawyers, provide critical legal expertise and practically support recourse for patients. 

\end{abstract}

\section{Introduction}
Many calls for AI accountability---accounting for, mitigating, and ultimately preventing harms from AI tools---in medicine focus on the interplay between physicians and patients.
Physicians are primarily identified as decision-makers, people that act alongside AI tools to make decisions, serving as buffers between AI recommendations and more vulnerable populations. Patients are primarily identified as decision-subjects, people directly subject to decisions made by an AI tool or someone using an AI tool.
The medical community has foregrounded the need for physicians' clinical judgment to provide human oversight over medical AI tools~\cite{NyrupAndRobinson2022, AbramoffEtAl2020, FroomkinEtAl2019}. 

Similarly, the legal and computer science communities have emphasized the need to design AI tools to support physicians' decision-making and prevent harms to patients~\cite{MansiEtAl2025_LegalXAI, PriceEtAl2019}.
Many call for Explainable AI (XAI) tools to provide human-understandable explanations for their reasoning and output~\cite{Lipton2018, ChariEtAl2020, GuidottiEtAl2018}. 
XAI tools are intended to enable physicians as decision-makers to prevent injury to patients and to advocate for a different course of action than recommended by an AI tool \cite{SunEtAl2025, CortiEtAl2024, ZajkacEtAl2025}. 
Explanations are intended to support people impacted by AI tools by enabling them to take pragmatic actions in response---that is, XAI tools aim to make medical AI tools actionable and contestable. Prior work on XAI tools for physicians focuses on integrating AI explanations into physicians' workflows to protect patients and support their health outcomes \cite{XieEtAl2020, GalsgaardEtAl2022,YangEtAl2019, CortiEtAl2024}, a theme echoed by the legal community who also emphasize XAI tools' role for determining fault for harms \cite{BaitAndSwitch2023, FroomkinEtAl2019, DesaiAndKroll2017}.

While there are incentives for companies to create AI tools that are actionable for decision-makers, they are not always equally incentivized to do so for decision-subjects \cite{MansiEtAl2025_LegalXAI}. 
Users, developers, and deployers of AI tools sometimes use the black-box nature of AI tools to evade liability, leaving decision-subjects without recourse~\cite{Statnews_Medicare, metcalf}.
For decision-subjects, contesting AI tools is fraught with power dynamics, particularly in contexts where decision-makers are not incentivized to advocate for decision-subjects. 
Examples of such contexts include housing~\cite{KuoEtAl2023}, health insurance claims processing~\cite{Review_Algorithm_2025}, or law enforcement~\cite{JinAndSalehi2024}. 
As a result, the actionability and contestability of an AI-based decision can greatly differ between patients and physicians, leaving patients vulnerable to others' decision-making.

In response to the challenges that decision-subjects face, the Human-Computer Interaction (HCI) Community has recognized the need to design technology to enable legal recourse for those harmed by AI-based decisions \cite{MansiEtAl2024_Contestability, Karusala2024Contestability, JinAndSalehi2024, WarrenAndSalehi2022}. If a patient is harmed from the use of a medical AI tool by a physician or medical insurer, they may seek a lawyer to pursue recourse for harms. Lawyers advocate for patients by integrating evidence from a case with legal information to construct a narrative that can persuade a judge or jury to act in favor of their client's perspective \cite{MansiEtAl2024_Contestability}, playing a critical role in the ``political process'' (i.e. power-related process) of contesting an AI decision in courts~\cite{Karusala2024Contestability}. Given the complex power dynamics between decision-makers, decision-subjects, and legal representatives, AI tools must account for the legal concerns of those impacted ~\cite{MansiEtAl2025_LegalXAI} to prevent negative consequences for the contestability of AI tools \cite{WarrenAndSalehi2022}. 

The HCI community has also recognized the impact of laws and regulations on how AI tools are created and used \cite{MansiEtAl2024_Contestability}, further impacting potential harms and recourse for patients. 
Laws and regulations impact how AI tools are designed, deployed, and developed. In response to challenges people are and will face with medical AI tools, the legal community is considering changes to how these tools are developed, including altering the kinds of datasets allowed for training AI models \cite{GerkeEtAl2020}; the requirements for evaluating AI tools before, during, and after deployment \cite{GerkeEtAl2020}, such as requirements for performance validation in localized settings \cite{AbramoffEtAl2020}; the standards for achieving FDA approval for AI tools as a medical device and software \cite{Chan2021, Price2022, FDA_AIMedical, Sundholm2024, GerkeEtAl2020}; and the responsibility of individual care providers in training and operating AI tools \cite{Chan2021, MalihaEtAl2021, Sundholm2024}. 

To better support AI accountability for patients, it is critical to inform changes to technological and legal infrastructures based on a direct and clear understanding of the challenges faced by patients and physicians. 
Effective communication and understanding around users' needs can help guide the creation of laws that have their intended effect on technology design and use; it also helps those creating AI tools to identify and implement methods for users to more easily pursue avenues for change when adverse outcomes happen \cite{MansiEtAl2024_Contestability}.

Analyzing current litigation and harms can identify patterns around how AI tools are impacting patient care and how to adapt technological and legal structures to better support AI accountability.
In our study, we conduct an analysis of AI litigation relative to different patient care tasks with which physicians engage. Based on litigation and reported harms, we first look to understand how AI tools impact physicians' medical tasks and result in harms to patients' outcomes. We then look to identify the implications of the current uses of AI tools for how stakeholders---including technologists, researchers, legal scholars, and policy makers---create more actionable AI tools in medical care.

In our analysis, we map out 31 legal cases and reported harms tied to AI tools that impact physicians' tasks in patient care. Our findings illustrate how AI tools used by stakeholders outside of the direct physician-patient relationship can significantly impact patients' care. We identify how many of these tools make decisions for vulnerable patient populations such as the elderly, those with physical disabilities, and those with mental health challenges. We describe how patients must often turn to legal representatives, such as lawyers or advocacy groups, to contest AI decisions in court and pursue the care prescribed by their own physicians. Based on our findings, we observe that XAI tools for physicians and other healthcare workers are important---but not in themselves sufficient---to reduce patient harms. Based on our findings, we identify legal and technological barriers to patients' abilities to seek recourse, and we describe considerations for addressing these barriers, identifying pathways for improving AI accountability.

\section{Results}
Because a large portion of conversations around AI accountability in medical settings center the impact of physicians as decision-makers, we looked to understand how harms from AI tools related to physicians’ patient care tasks. 
We adopted a case study approach to thematically analyze real-world cases of reported harms and litigation for AI tools impacting doctors' tasks in patient care. 

To identify legal cases, we manually reviewed entries in two databases, the Database of AI Litigation (DAIL) \cite{DAIL} and the Health Litigation Tracker \cite{Georgetown_LitigationTracker}, both known to track health care litigation in the United States including lawsuits involving AI tools in healthcare delivery. The final case set included  12 entries about legal cases, which we will refer to in this paper with the following identifiers used in the two litigation trackers (DAIL: Dinerstein v. Google, LLC; Ark. Dep’t of Human Servs. v. Ledgerwood; K.W. v. Armstrong; Michael T. v. Bowling; C.S. et al v. Saiki; Barrows et al v. Humana, Inc.; Estate of Gene B. Lokken et al. v. United-Health Group, Inc. et al.; Kisting-Leung v. Cigna Corp.; Van Pelt v. Cigna Group; O’Neill: Humana Inc. et al. v. Kennedy et al.; Estate of Gene B. Lokken et al. v. UnitedHealth Group, Inc. et al.; Kisting-Leung v. Cigna Corp.). 

Because significant harms may not be litigated, we also manually searched for reported harms through the AIAAIC Database, a public interest initiative that collects and classifies issues driven by and relating to AI, algorithms, and automation. The AIAAIC is regularly cited and mentioned by academic, non-profit, advocacy, government, and commercial research institutes, associations, and agencies \cite{AIAAIC_Mentions}. This resulted in a final case set of 19 entries about harms, which we will refer to in this paper with the following identifiers used in the AIAAIC (AIAAIC0106; AIAAIC0159; AIAAIC0316; AIAAIC0633; AIAAIC1272; AIAAIC0406; AIAAIC1509; AIAAIC0758; AIAAIC0668; AIAAIC1785; AIAAIC0647; AIAAIC1251; AIAAIC1841; AIAAIC1252; AIAAIC1720; AIAAIC1900; AIAAIC1061; AIAAIC1838; AIAAIC0248). For full details on how we accessed and reference entries in all three databases see the ``Data Availability'' section in the Methods.

We identified 31 case studies that met our criteria: (a)~if patient care was or could be impacted, (b)~if the harm impacts a task physicians normally take in patient care; and (c)~if the harm or legal case occurred in the United States (we only considered entries in the United States because international laws and litigation processes can differ greatly). For full details on eligibility criteria and selected cases, see the ``Eligibility Criteria'' and ``Data Selection'' sections in the Methods. 

\begin{figure}[t]
  \includegraphics[width=0.48\textwidth]{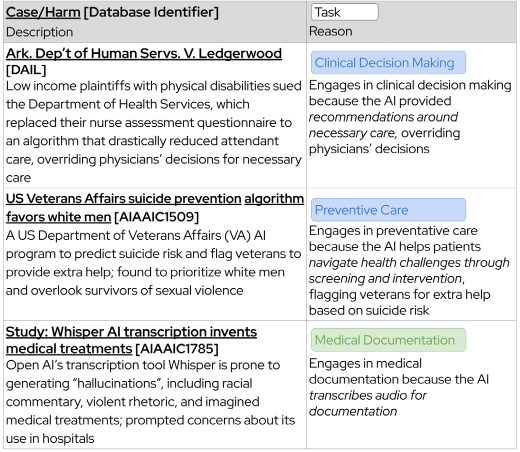}
  \caption{This figure shows three cases used in our case study analysis of harms or litigation involving AI technologies in medical settings. In our analysis cases are organized according to physician task (colored box) along with a summary reason for the classification.}
 \label{fig:ExampleCaseCoding}
\end{figure}

The research team then used an iterative approach to identify and code each legal case or harm with the main physician task(s) impacted. Cases and harms were coded based on a list of 10 definitions and examples of tasks physicians undertake when caring for patients, which the research team co-created with a collaborating physician. Three example cases are given in Figure~\ref{fig:ExampleCaseCoding} showing the reference to the case, a description of the harm, the physician task label, and the reason for the choice of label. See the ``Data Analysis'' section in the Methods for more information about physician tasks used and how each entry was coded. See Supplementary Figures 1-3 for the full list of all legal cases and harms we tracked, including the name of the case, the reference to the citation in our paper, the physician task it corresponds to, and the reason for categorizing it under certain tasks. For legal cases, we also note which cases are ongoing.

We present our findings, which we then use to discuss how AI tools used by stakeholders outside of the direct physician-patient relationship can significantly impact patients' care.

\subsection{Patterns Across Clinical Decision Making}

\begin{figure}[t]
  \includegraphics[width=0.47\textwidth]{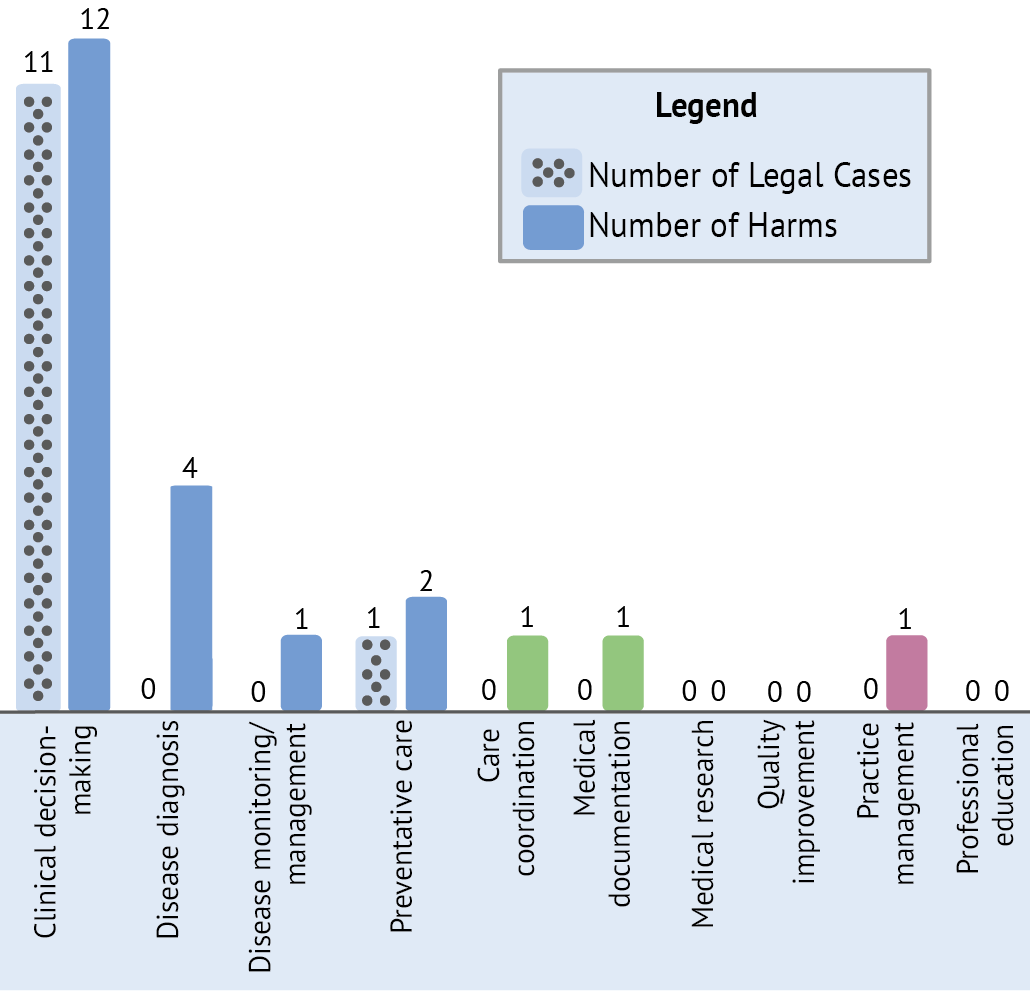}
  \caption{The distribution of legal cases and harms across physicians' tasks. Most legal cases and harms centered on AI tools that infringe on physicians' clinical decision making.}
 \label{fig:Counts}
\end{figure}

Most legal cases and harms centered on AI tools that impact physicians' Clinical Decision Making tasks. Of the 12 legal cases we identified, 11 involved an AI tool impacting Clinical Decision Making. The other legal case was tied to Preventive Care. Clinical Decision Making was also the physician task associated with the most harms (12); followed by Disease Diagnosis (4); Preventive Care (2); Practice Management, Care Coordination, Medical Documentation, and Disease Monitoring/Management (1). No harms or legal cases in our review were tied to AI tools involved in Medical Research, Quality Improvement, or Professional Education. Figure~\ref{fig:Counts} shows the distribution.

Most of the harms and legal cases tied to Clinical Decision Making originated from the use of AI tools by non-physician stakeholders. 
Only three of the 12 harms associated with Clinical Decision Making were tied to algorithms intended to directly interact or interface with physicians. IBM's Oncology Expert Advisor aimed to provide personalized cancer treatment recommendations for physicians (AIAAIC0106; AIAAIC0159). 
Harms to patients were reported around a Google Cloud (AIAAIC0647)
collaboration to create algorithms for patient monitoring and medical decision making.

In contrast, the other nine harms and all 11 legal cases tied to Clinical Decision Making involved the use of AI tools for determinations by non-physician entities. For example, several prominent lawsuits involved U.S. state governments 
(DAIL: Ark. Dep’t of Human Servs. v. Ledgerwood;
Michael T. v. Bowling; K.W. v. Armstrong; C.S. et al v. Saiki; AIAAIC1720) that used AI tools to change the allocation of health benefits, significantly cutting resources.
Most of the other harms and legal cases reported involved health insurance companies that used AI tools to deny claims, hindering patients from medically necessary treatments 
(DAIL: Barrows et al v. Humana, Inc.; Estate of Gene B. Lokken et al. v. UnitedHealth Group, Inc. et al.; Van Pelt v. Cigna Group; Kisting-Leung v. Cigna Corp.; O’Neill: Humana Inc. et al. v. Kennedy et al.; Estate of Gene B. Lokken et al. v. UnitedHealth Group, Inc. et al.; Kisting-Leung v. Cigna Corp.; AIAAIC1252; AIAAIC1841;
AIAAIC1900; AIAAIC1061; AIAAIC1838; AIAAIC0248;
AIAAIC1251).

AI tools used by other stakeholders do not involve the patient's physician, but they make AI determinations that contradict or override physicians' clinical decision making and create negative patient outcomes. Several prominent legal cases 
(DAIL: Barrows et al v. Humana, Inc.; Humana Inc. et al. v. Kennedy et al.; Van Pelt v. Cigna Group; O’Neill: Humana Inc. et al. v. Kennedy et al.) involve health insurance companies that denied medically necessary care to elderly patients based on an AI prediction of patient recovery times. Even when physicians and medical documentation clearly indicated a patient needed much longer to recover, the decisions made by these AI tools overruled the physician determination and patients' claims were denied. Consequently, these systems effectively provided clinical decisions that overrode the decisions of patients' physicians and materially impacted their medical care. In several harms 
(AIAAIC0758; AIAAIC1838) companies, under the pretense of preventing fraud, used AI tools to make financially profitable health determinations that negatively impacted patients without a physician's review.

\subsection{A Snapshot of People Impacted}
Across all harms and litigation, most AI tools made determinations primarily for vulnerable populations. Vulnerable populations included the elderly 
(DAIL: Barrows et al v. Humana, Inc.; Estate of Gene B. Lokken et al. v. UnitedHealth Group, Inc. et al.; O’Neill: Estate of Gene B. Lokken et al. v. UnitedHealth Group, Inc. et al.; AIAAIC1252; AIAAIC1841),
those with physical disabilities (DAIL: Ark. Dep’t of Human Servs. v. Ledgerwood; K.W. v. Armstrong; Michael T. v. Bowling; C.S. et al v. Saiki; AIAAIC1720), and those with mental health challenges (AIAAIC1838). No harms or legal cases involved claims against physicians for decisions made with an AI tool.

Lawsuits were undertaken by either private individuals impacted by the system or by advocacy groups, such as the ACLU. Cases involving AI tools used by governments, such as those used by states, were often stopped more quickly because determinations impacting vulnerable populations violated their rights. State judges ordered injunctions, which require the state to immediately stop using the system, and ordered states to restore use of their previous algorithms for resource allocations. Algorithms deployed by private companies were deployed for longer and often stopped when private individuals sued companies.

\subsection{Patterns Across Physicians' Other Tasks}
Poorly functioning systems deployed to physicians were retracted. Most AI harms and litigation outside of Clinical Decision Making involved systems deployed for use by physicians. Our review included a total of 11 harms and legal cases in Monitoring/Management, Disease Diagnosis, Preventive Care, Care Coordination, Medical Documentation, and Practice Management. 
Notably, just under half of these harms (5 of 11) involved systems intended for use by physicians for tasks other than Clinical Decision Making, but most of these systems were retracted in response to complaints of poor performance.

\section{Discussion}

The harms and legal cases we reviewed point to how patients’ medical care relies on a complex web of stakeholders, all of which play a role in healthcare delivery and contribute to patient outcomes. 
Many conversations around XAI tools and AI accountability are physician-centered---they focus on tailoring AI tools for physicians' medical workflows and needs to protect themselves and their patients \cite{XieEtAl2020, GalsgaardEtAl2022, YangEtAl2019, CortiEtAl2024}.
However, as described in our Findings,
AI tools outside of the direct physician-patient relationship can deeply impact patients’ care.

Additionally, patients experience significant barriers when contesting and seeking recourse against harms from non-physicians using AI tools in decision making. Users, developers, and deployers of AI tools can use the black-box nature of systems to evade liability~\cite{Statnews_Medicare, metcalf}. As we described, patients had to pursue lawsuits to contest AI-based decisions. 
Many lawsuits we reviewed were class action lawsuits, reflecting repeated errors and harms across individuals. 
Those that were not initiated by individuals were undertaken by public advocacy groups, pointing to collective public harm and significant barriers to seeking recourse by individual patients. 

\begin{figure*}[t]
  \centering
 \fbox{\includegraphics[width=0.95\linewidth]{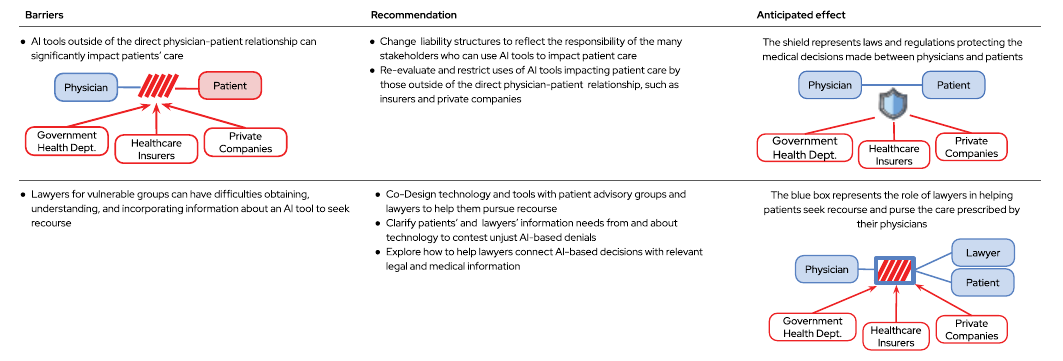}}
 \caption{In this figure, we summarize our findings and recommendations for the legal and HCI communities to help address patients' barriers to recourse. For the legal community, we recommend changes to liability structures to reflect the responsibility of the many stakeholders who can use AI tools to impact patient care. For the HCI community, we recommend co-designing with patient advisory groups, lawyers, and patients to create technology to support recourse.}
 \label{fig:Considerations}
\end{figure*}

Re-framing AI accountability in response to a patient-centered perspective of harms brings to light opportunities for creating AI tools, explanations, and legal structures that support contestability by patients. 
Because of the joint role of laws and regulations and technology design in supporting AI accountability, we describe opportunities for the legal and HCI communities to help patients push back against AI-based decisions. Figure~\ref{fig:Considerations} summarizes the legal and technological considerations for redress that we discuss.
 
The first legal barrier our study identifies relates to how the law delegates responsibility for harms from the use of AI tools. The complex web of people and their responsibilities is not fully reflected in the law when it comes to harms and accountability from the use of AI tools impacting medical decisions.
As seen from our analysis, physicians are only partially responsible for patient outcomes, relative to the impacts of other actors.
Healthcare insurers, national and state health departments, and private companies are impacting patient care and monitoring---from outside of the direct physician-patient relationship.
For decades, scholars have warned against the commodification of medicine as a trend that fractures the relationships between physicians and their patients \cite{Pellegrino1999}, and AI tools may exacerbate health insurers' and private companies' existing financial incentives.
Physicians---who are only one group of many actors shaping AI tools in patient care---have good reason to be concerned about current liability structures as they relate to the deployment of AI tools and the safety of their patients \cite{PriceEtAl2019}. 
Without changing either financial incentives or legally permitted applications, patients are at risk and will continue to bear an unfair burden of harms and challenges when contesting AI-based decisions across healthcare delivery.

Reassessing the legal structures that contribute to deterrence and compensation from a patient-centered perspective can better align AI accountability with the landscape of harms and deter harmful AI deployment.  
Legal scholars~\cite{Sundholm2024, Price2022, Chan2021} are proposing distributed models of risk where physicians, AI creators, hospital administrators, and other stakeholders share legal responsibility when patients are harmed from the use of AI tools. 
Changing legal responsibility structures to reflect the aggregate responsibility of stakeholders using AI tools may better deter companies or administrators from deploying AI tools and more equally reflect their shared responsibility in how patients are compensated when they are harmed. 

Re-evaluating the acceptable uses of AI tools by insurers and private companies can also better protect patients and their physicians.
Evaluating existing laws and regulatory structures in light of patient harms and vulnerabilities can guide discussions around incentives and guidelines, such as granting the FDA more authority and sufficient resources to review applications of AI tools \cite{Olivia2025, Review_Algorithm_2025} and banning AI tools for medical insurance decisions impacting patient care.

The first technological barrier our study identifies relates to shifting the focus of AI tools and explanations to scaffolding patients' paths to recourse. 
Our analysis highlights how only focusing on AI tools for physicians obscures where harms are befalling patients, creating an illusion of AI accountability without preventing harms or helping patients obtain recourse.
Poorly functioning AI tools deployed to physicians were largely retracted in response to users' complaints. However, when companies deployed systems that disrupt the relationship between patients and their physicians, patients had to seek legal recourse. Consequently, AI tools for healthcare workers---including physicians, social workers, and nurses---are important, but not sufficient, to protect and support patients.

Our examination of the data verifies what others~\cite{MansiEtAl2024_Contestability, alfrink2023contestable} have postulated about AI tools impacting medical care: the critical role of legal representatives in AI accountability. 
Prior work designing for stakeholders' legal needs has focused on topics tied to physicians' legal concerns, such as those around malpractice liability \cite{AbramoffEtAl2020, PriceEtAl2019}. In other settings, access to navigational support, including legal expertise, can help marginalized groups engage with bureaucracies, construct effective appeals, and gain insight into AI tools across multiple decisions~\cite{Karusala2024Contestability}.

Designing actionable AI tools for patients' legal representatives is critical for vulnerable patients to seek recourse against powerful stakeholders, such as healthcare insurers, in the courts. Algorithmic solutions for AI accountability need to evaluate if and how their explanations will be actionable for patients and their legal representatives. 
Defendants' lawyers may face challenges obtaining even basic information about an AI tool and how it may impact their clients \cite{JinAndSalehi2024}.
Companies or other decision-makers are dis-incentivized to convey information that would be to the benefit of the decision subject and to the disadvantage of the decision maker.
Information may be presented in the form of dark patterns---explanations intentionally designed to discourage actionability~\cite{chromik2019darkXAI}---or explainability pitfalls---explanations that unintentionally have that effect~\cite{Ehsan2021Pitfalls}. Additionally, lawyers can encounter barriers to building cases against companies that deploy AI tools despite access to technical performance evaluations of said systems or explanations of their behavior; and they also face difficulties understanding how these technologies are developed and used~\cite{JinAndSalehi2024}.

To empower patients and their lawyers to seek recourse, AI tools should help them understand the sociotechnical context impacting AI-mediated decisions.
AI tools in healthcare are embedded in complex sociotechnical environments with people who may have conflicting roles and relationships to patients. 
Consequently, AI tools must be designed considering lawyers' needs to incorporate different human and algorithmic aspects of an AI system's sociotechnical context \cite{SunEtAl2022, Ehsan2020HumancenteredEA, Ehsan2021Pitfalls, EhsanEtAl2024_HCXAI-LLM, HarmanpreetEtAl2022, ChenEtAl2025}. 
For example, it may be helpful to understand how algorithmic aspects of an AI tool's sociotechnical context should be provided alongside information about whether others have chosen to accept or contest AI decisions in the past, why, and how the decision was contested. This mirrors the approach of prior human-centered XAI work positing a pluralist definition of explanations, reflecting how different groups vary in their interpretations of technological artifacts \cite{CortiEtAl2024, EhsanEtAl2023_ComingofAge}.

\section{Next Steps}
It is important to work alongside patients, patient advisory groups, and lawyers to address technological and legal barriers to recourse. Co-design approaches can provide space for experts in diverse fields to contribute and share knowledge \cite{VinesEtAl2013}, supporting the cross-disciplinary communication central to improving contestability.
Co-design with these stakeholders can advance patient-centered notions of AI accountability in several ways.

First, co-design can help address legal barriers by supporting cross-disciplinary communication between the legal and HCI communities. Cross-disciplinary communication is critical for improving laws supporting more contestable AI tools. 
Research has recognized the importance of laws in enforcing ethical and safe designs \cite{alfrink2023contestable}, and the joint role of lawyers and other AI creators in supporting AI accountability \cite{OtterloAndAtzmueller2018, alfrink2023contestable, Oswald2013, MansiEtAl2024_Contestability}. 
Co-design methods can help ensure laws have their intended effect on technology design. For example, some have highlighted the role that speculative co-design practices could play in understanding the potential impact of technologies and discussing the policy dimensions of their implementation \cite{SpaaEtAl2019, WarrenAndSalehi2022}.

Co-design can help address technological barriers in several ways as well. Co-design with patient advisory groups and lawyers can uncover their information needs for contesting unjust AI-based decisions, and support cross-disciplinary communication with developers. In this way, developers may better design AI tools' technical specifications to support contestability.
For example, co-design methods could help with tailoring AI explanations for requirements around contestability \cite{LeofanteEtAl2024}, ensuring AI tools have the legally-relevant information about an underlying algorithm.  

Further, co-design methods can pinpoint areas of innovation beyond explanations. Prior work has shown how challenges working with AI tools can undermine lawyers' legal defense, translating to significant consequences for clients \cite{WarrenAndSalehi2022}. Consequently, there's a need to address barriers in engaging lawyers in ``evaluation design and building tools'' \cite{JinAndSalehi2024}. Co-design methods with lawyers could clarify the role and design of AI-based tools, including identifying needs for sociotechnical gaps that AI tools and explanations cannot meet.

By understanding the harms patients are experiencing from AI tools, we can identify and expand their pathways to contesting unfair or incorrect AI-based decisions. 
As our analysis shows, AI tools are used outside of the direct patient and physician relationship in ways that disrupt and negatively impact patients' health outcomes, leaving them no choice but to seek legal recourse. 
By understanding how patients are contesting AI tools, we can expand and strengthen existing actionability pathways, bolstering their abilities to pursue medical care.

\section{Acknowledgments}
\section{Funding Statement}
The author gratefully acknowledges  the support of the National Science Foundation GRFP under Grant No. DGE-2039655. Any opinion, findings, and conclusions or recommendations expressed in this material are those of the authors(s) and do not necessarily reflect the views of the National Science Foundation. 

\section{Author Contributions Statement}
Conceptualization: GM (lead), MR (equal) \\
Data curation: GM (lead), MR (supporting) \\
Investigation: GM (lead), MR (supporting) \\
Methodology: GM (lead), MR (equal) \\
Supervision: MR (lead), GM (supporting) \\
Validation: GM (lead), MR (equal) \\
Visualization: GM (lead), MR (equal) \\
Writing – original draft: GM (lead), MR (equal) \\
Writing – review \& editing: GM (lead), MR (equal)

\section{Competing Interests Statement}
The authors declare there are no competing interests.

\section{Figure Legends/Captions (for main text figures)}
\textbf{Figure 1:} This figure shows three cases used in our case study analysis of harms or litigation involving AI technologies in medical settings. In our analysis cases are organized according to physician task (colored box) along with a summary reason for the classification. \\

\noindent \textbf{Figure 2:} The distribution of legal cases and harms across physicians’ tasks. Most legal cases and harms centered on AI tools that infringe on physicians’ clinical decision making. \\

\noindent \textbf{Figure 3:} In this figure, we summarize our findings and recommendations for the legal and HCI communities to help address patients’ barriers to recourse. For the legal community, we recommend changes to liability structures to reflect the responsibility of the many stakeholders who can use AI tools to impact patient care. For the HCI community, we recommend co-designing with patient advisory groups, lawyers, and patients to create technology to support recourse.

\bibliographystyle{plain}

\section{Methods}
Because a large portion of conversations around AI accountability in medical settings center the impact of physicians as decision-makers, we looked to understand how harms from AI tools related to physicians’ patient care tasks. 
We adopted a case study approach to thematically analyze real-world cases of reported harms and litigation for AI tools impacting doctors' tasks in patient care. The legal cases and harms identified and reviewed in this study are not intended to be exhaustive or representative of all potential harms that patients may experience from medical AI tools; rather, they were selected to illustrate the range and nature of publicly documented issues.

\subsection{Search Strategy}
To identify legal cases, we reviewed entries in two databases, the Database of AI Litigation (DAIL) \cite{DAIL} and the Health Litigation Tracker \cite{Georgetown_LitigationTracker}, both known to track health care litigation in the United States including lawsuits involving AI tools in healthcare delivery. Because significant harms may not be litigated, we also searched for reported harms through the AIAAIC Database, a public interest initiative that collects and classifies issues driven by and relating to AI, algorithms, and automation. The AIAAIC is regularly cited and mentioned by academic, non-profit, advocacy, government, and commercial research institutes, associations, and agencies \cite{AIAAIC_Mentions}.

\subsection{Eligibility Criteria}
We identified 31 case studies that met our criteria: (a)~if patient care was or could be impacted, (b)~if the harm impacts a task physicians normally take in patient care; and (c)~if the harm or legal case occurred in the United States (we only considered entries in the United States because international laws and litigation processes can differ greatly). 

In the U.S., physicians are legally responsible for protecting patient privacy, so we included entries related to privacy cases. We also included entries where non-physicians use AI tools to impact physicians' tasks. For example, we include cases where insurance companies use benefits payments to deny care for patients because these tools provide recommendations about necessary care that effectively override physicians' clinical decision making, resulting in a harm impacting a task that physicians normally take.

Entries tied to the deployment of AI tools with social workers and nurses \cite{AIAAIC_Oregon, AIAAIC_HeartAttacks}, 
were excluded either because they did not take place in the U.S. or they did not impact physicians' medical decision making in a clinical setting. While their roles in patient care and needs from AI tools are important, we restricted the scope of our current study to all cases impacting patient care by U.S. physicians.

\subsection{Data Selection}
In February 2025, the first author manually reviewed all entries in the AIAAIC database resulted in the initial selection 2207 entries reported to occur between 2016-2024. The first author screened for relevancy by reviewing the titles of each entry, leaving a total of 93 entries for further review. Each remaining entry was screened for the eligibility criteria based on the paragraph summary that each entry has. Additional news articles or other links on a database entry were reviewed to clarify impacts to patient care or impacts on clinical task as needed.

This resulted in a final case set of 19 entries about harms (AIAAIC0106; AIAAIC0159; AIAAIC0316; AIAAIC0633; AIAAIC1272; AIAAIC0406; AIAAIC1509; AIAAIC0758; AIAAIC0668; AIAAIC1785; AIAAIC0647; AIAAIC1251; AIAAIC1841; AIAAIC1252; AIAAIC1720; AIAAIC1900; AIAAIC1061; AIAAIC1838; AIAAIC0248).

To search the O'Neill Institute's Health Litigation tracker, the first author manually reviewed 195 entries that take place from 2017-2024. The first author screened for relevancy by reviewing the ``Why this Matters'' and ``Potential Impact'' summaries of each entry, referencing the full entry to clarify impacts to patient care or impacts on clinical tasks as needed.

To search DAIL, the first used the ``Applications'' field to identify potential, relevant cases. Each entry in DAIL is tagged with at least one of 48 ``Applications'' fields. Based on the range of harms and AI tools documented in the AIAAIC, the first author used combinations of the following nine options for the ``Applications'' field to search widely for potential, relevant legal cases: Biometrics Data, Disabilities Benefits, DNA Analysis, Fraud, Generative AI, Health, Agentic AI, Privacy, Public Benefits. This search yielded 171 entries for legal cases from DAIL. The first author screened for relevancy by reviewing the ``Brief Description'' field of each database entry. If needed, entries were expanded to view additional legal documents or primary sources to identify if and how the cases met our criteria.

The final case set included  12 entries about legal cases (DAIL: Dinerstein v. Google, LLC; Ark. Dep’t of Human Servs. v. Ledgerwood; K.W. v. Armstrong; Michael T. v. Bowling; C.S. et al v. Saiki; Barrows et al v. Humana, Inc.; Estate of Gene B. Lokken et al. v. United-Health Group, Inc. et al.; Kisting-Leung v. Cigna Corp.; Van Pelt v. Cigna Group; O’Neill: Humana Inc. et al. v. Kennedy et al.; Estate of Gene B. Lokken et al. v. UnitedHealth Group, Inc. et al.; Kisting-Leung v. Cigna Corp.)

The tallies of legal cases and entries about harms are absolute numbers of the total cases we found across all databases, including entries from different databases mentioning the same legal case. There were 5 entries that covered overlapping legal cases across databases, and 3 harms entries that overlapped. Because AIAAIC and the litigation trackers are living databases---actively updated and maintained by the community---they could contain different information. Further, they could cover different aspects of harm presented to patients. Consequently, we opted to keep and cite all duplicates independently, rather than selecting one.

\subsection{Data Analysis}
The first author manually all entries selected. For both legal databases, the first author also reviewed public sources about each case, including legal documents, and other primary sources. If it was a legal case, the first author noted the general legal issue in question (e.g. malpractice, privacy), the primary defendants and prosecutors, and, if available, other complaints or concerns outside of the issue represented in the litigation. 
If it was a harm, the first author noted the general issue for which the tool was criticized, and the primary concern or harm from the system's application. 

\begin{figure}[h!]
  \includegraphics[width=0.45\textwidth]{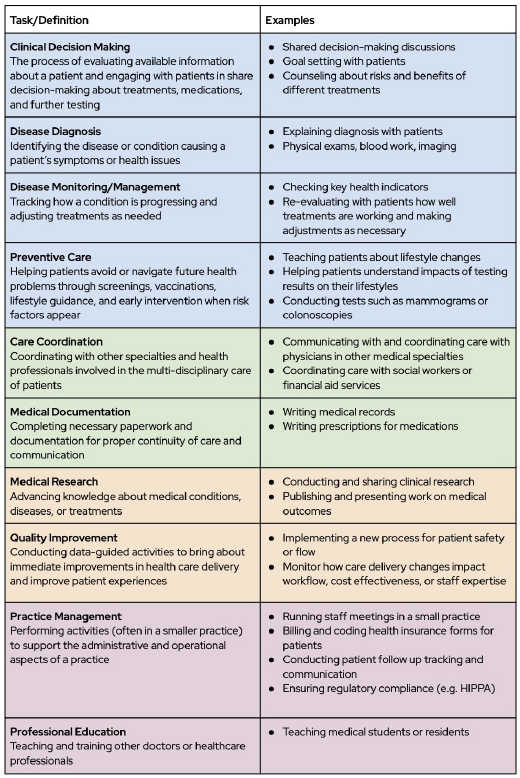}
  \caption{The physicians' tasks, definitions, and examples that we created and referenced as we analyzed legal cases and harms. Tasks at the top of the list are more patient-facing and are increasingly less patient-facing further down.}
 \label{fig:DoctorsTasks}
\end{figure}

The research team then used an iterative approach to identify and code each legal case or harm with the main physician task(s) impacted. Cases and harms were coded based on a list of 10 definitions and examples of tasks physicians undertake when caring for patients, which the research team co-created with a collaborating physician. Tasks ranged from more
patient-facing, such as clinical decision-making and disease monitoring, to less, such as professional education and practice management. For each task, we created a set of example
physician activities that could be used to further guide our analysis (see Figure~\ref{fig:DoctorsTasks}). If a case or harm overlapped with more than one physician task, we coded it with all related tasks.

\subsection{Data Availability}
As of February 2025, the three databases we referenced are open and free to access online. Below we provide details about how to access the legal cases and entries we reference in this paper. Because AIAAIC and the litigation trackers are living databases---actively updated and maintained by the community---they could contain different information than what we reviewed in February 2025. 

\subsection{Code Availability}
No code was used to produce the results of the study.

\subsubsection{Referencing and Accessing Entries from DAIL and the O'Neill Institute. }
We reference legal cases and entries through unique identifiers in each database. For DAIL and the O'Neill Institute, we reference the case names, which can be used as a keyword to locate the entries in these databases. 

\subsubsection{Referencing and Accessing Entries from the AIAAIC. }
The AIAAIC has multiple kinds of databases. The ``Incidents Database'' tracks reported incidents of harms; The ``Systems Database'' summarizes ethical issues for specific AI systems and contains links to entries in the Incidents Database. For example, the Systems Dabase has an entry for the nH Predict system and contains links to each of the entries in the Incidents Database that are related to the nH Predict system. To avoid duplicate entries and to center harms rather than systems, we included the pages from the Incidents Database, but excluded the entries from the Systems Database.

For the AIAAIC, we use the AIAAIC ID number, used to track entries in the AIAAIC's Incidents Database. To access the Incidents database, begin at the \href{AIAAIC}{https://www.aiaaic.org} main page. Navigate to the ``AIAAIC repository'' tab and select ``AI, Algorithmic, and Automation Incidents'' from the drop down. Then click on ``Access Database'', which leads to a Google Sheet. Select the ``Incidents'' tab. The first column of the spreadsheet says ``AIAAIC ID\#'', which match the ID's listed in this publication.

\subsubsection{Included Legal Cases and Harms}
Below we list all 31 of the legal cases and harms that we included in our study, along with our reasoning for categorizing it as a specific physician task.
\begin{figure}[h!]
  \includegraphics[width=0.47\textwidth]{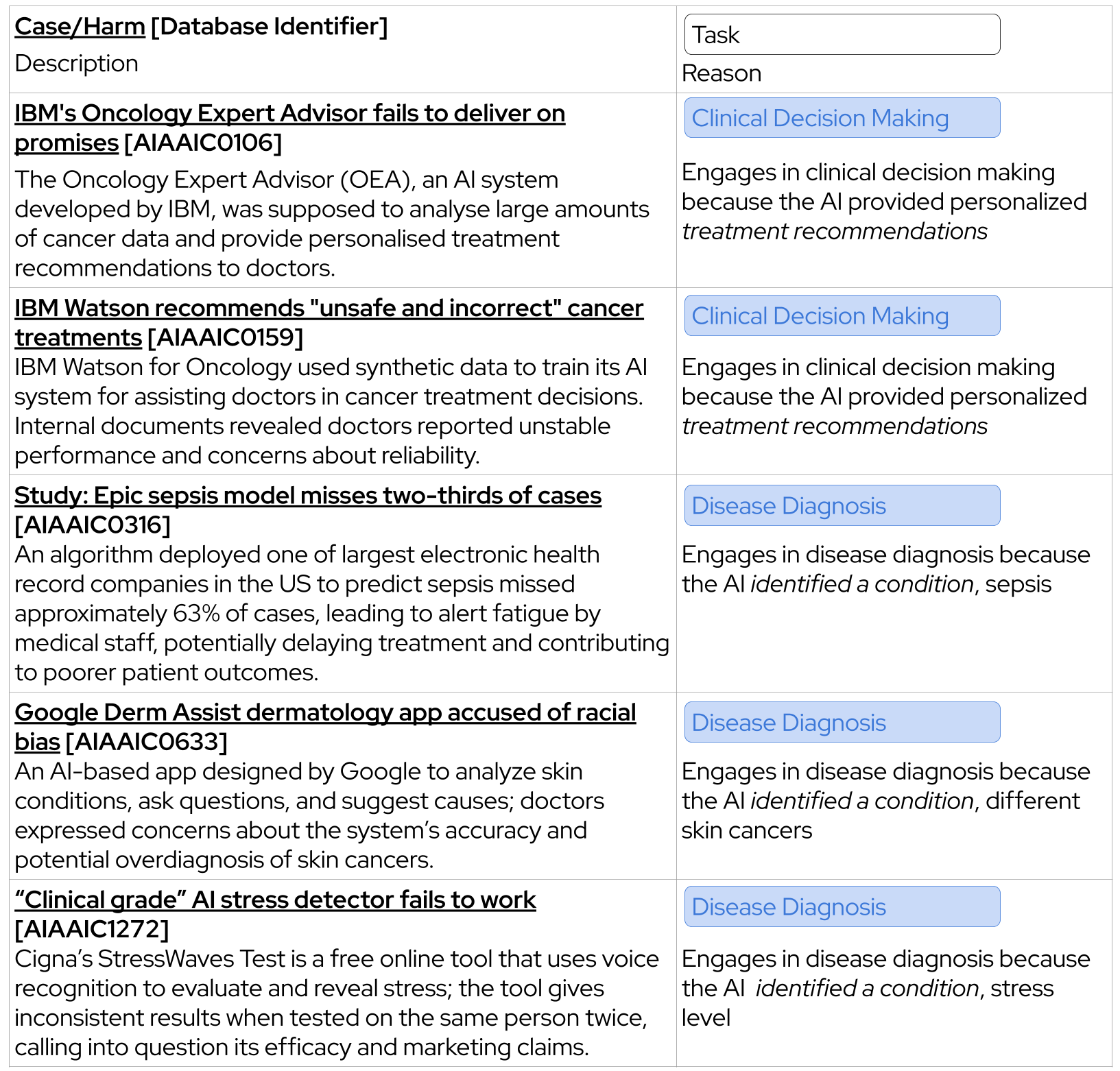}
  \includegraphics[width=0.47\textwidth]{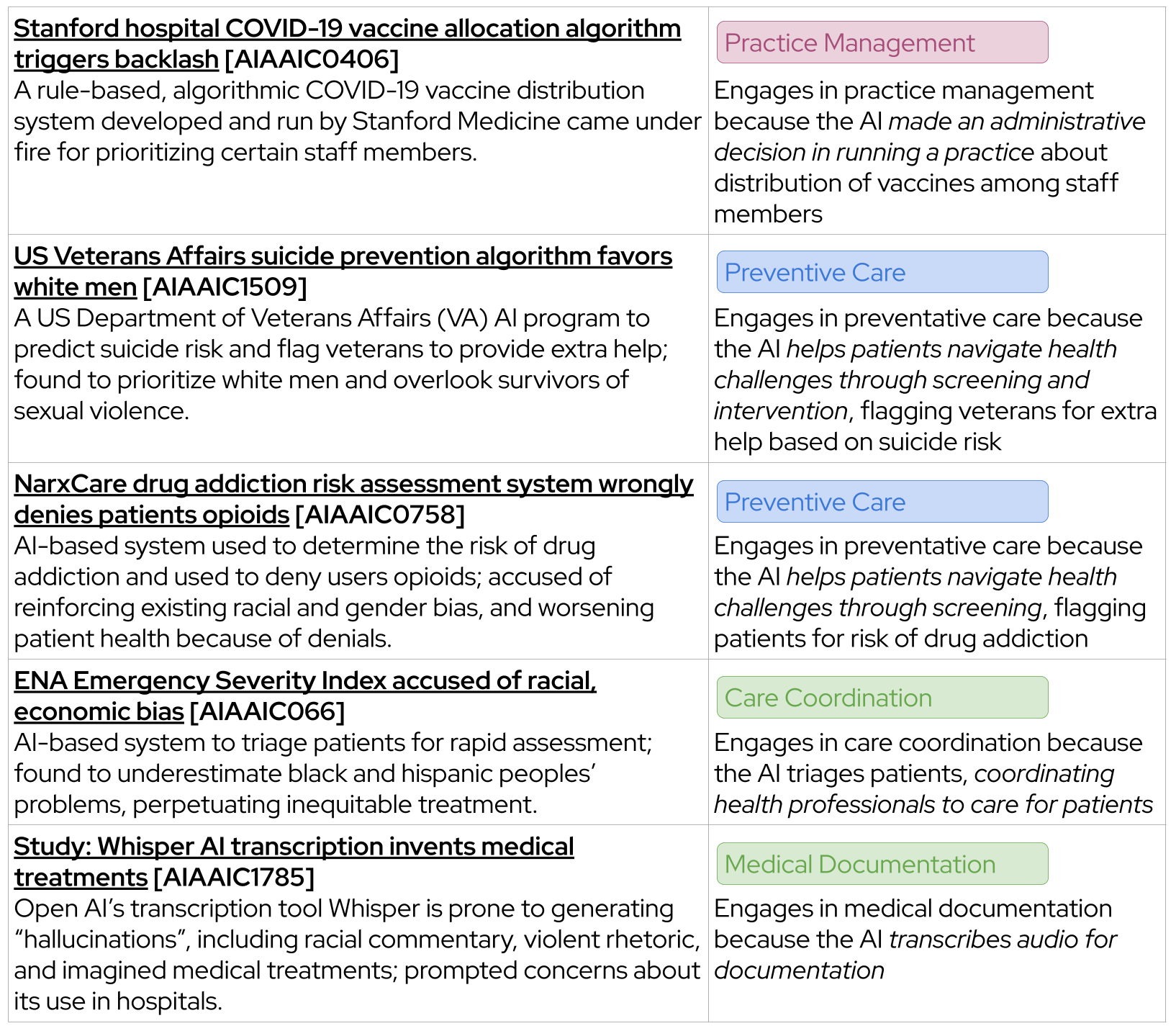}
  \caption{The legal cases and harms we tracked in our case study. We include the name of the case, the reference to the citation in our paper, the physician task it corresponds to, and the reason for categorizing it under certain tasks. For legal cases, we note which cases are ongoing.}
 \label{fig:LegalCases-Harms_1}
\end{figure}

\clearpage

\begin{figure}[t]
  \vspace{-15ex}
  \includegraphics[width=0.47\textwidth]{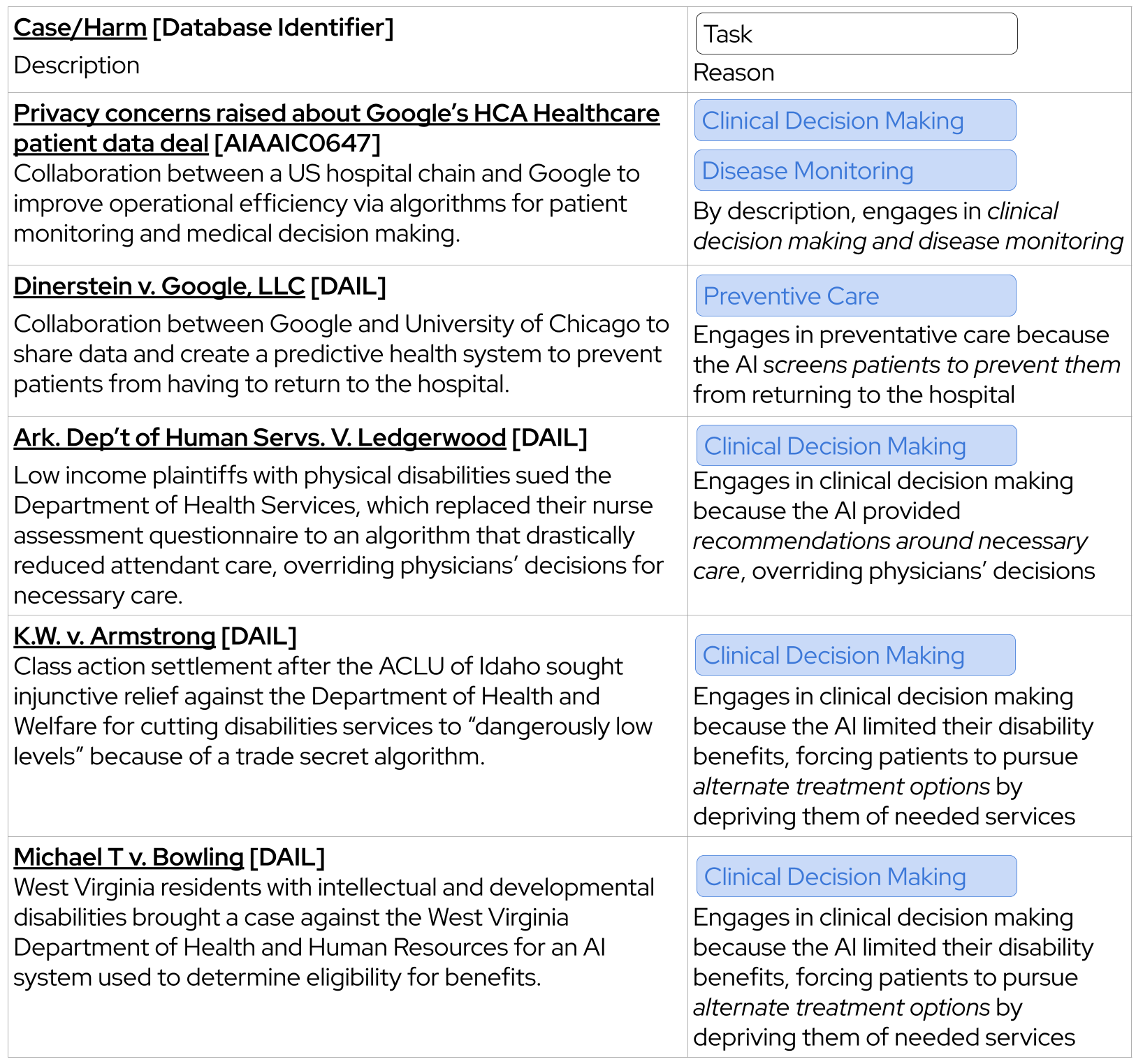} \hfill
  \includegraphics[width=0.474\textwidth]{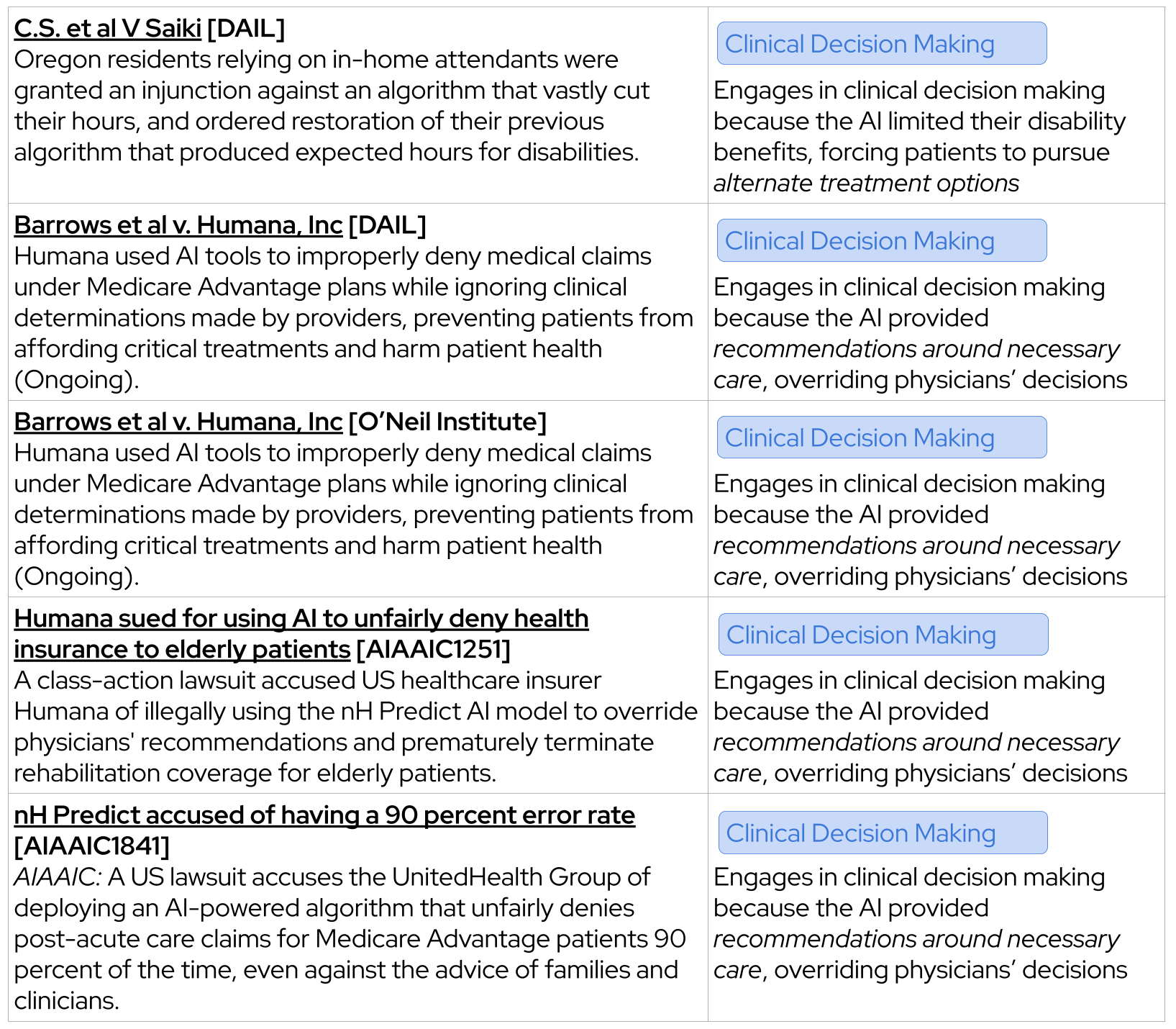} \hfill
  \includegraphics[width=0.473\textwidth]{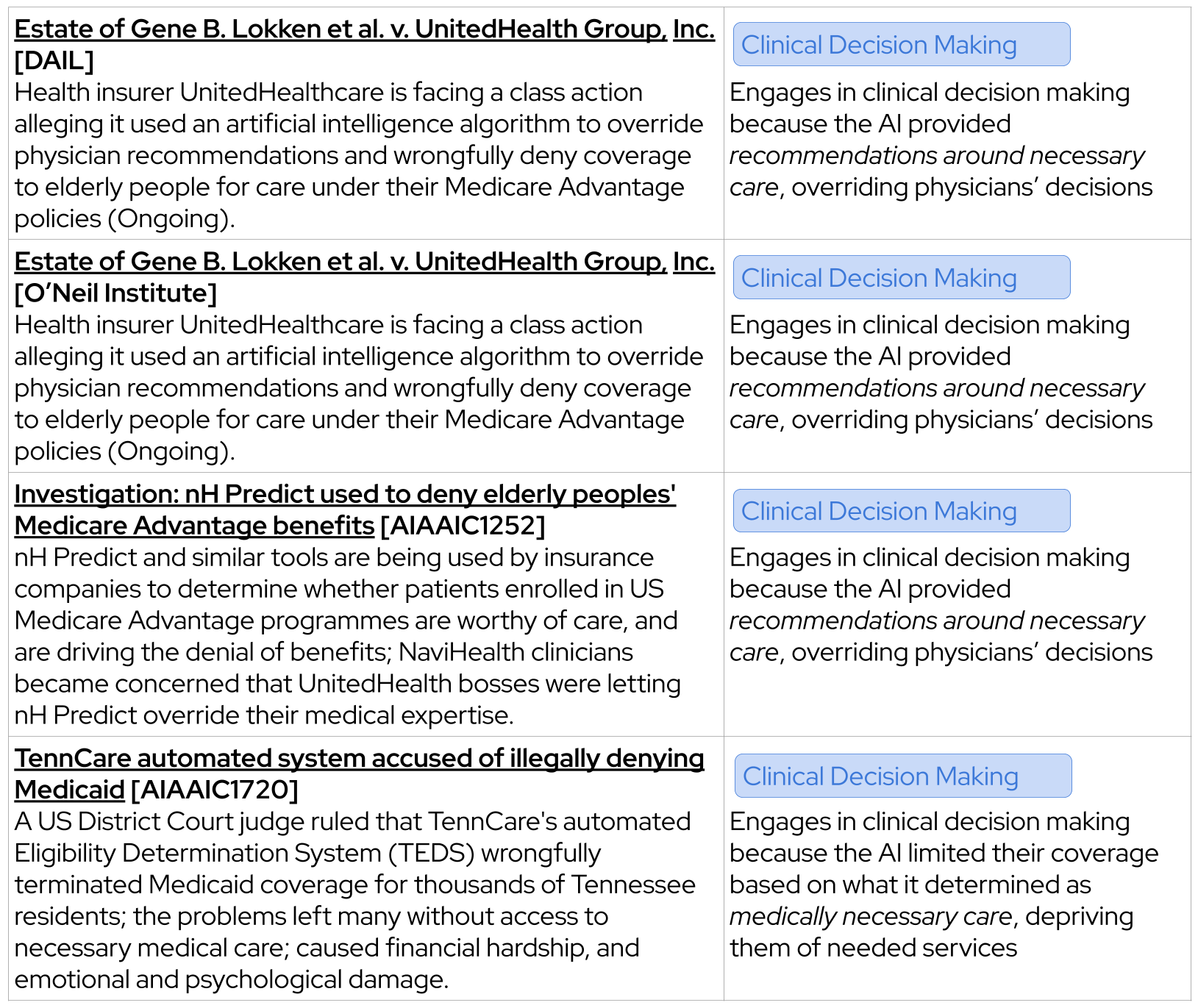} \hfill

  \caption{(continued) The legal cases and harms we tracked in our case study.}
 \label{fig:LegalCases-Harms_2}
\end{figure}

\begin{figure}[t]
  \vspace{-74ex}
  \includegraphics[width=0.47\textwidth]{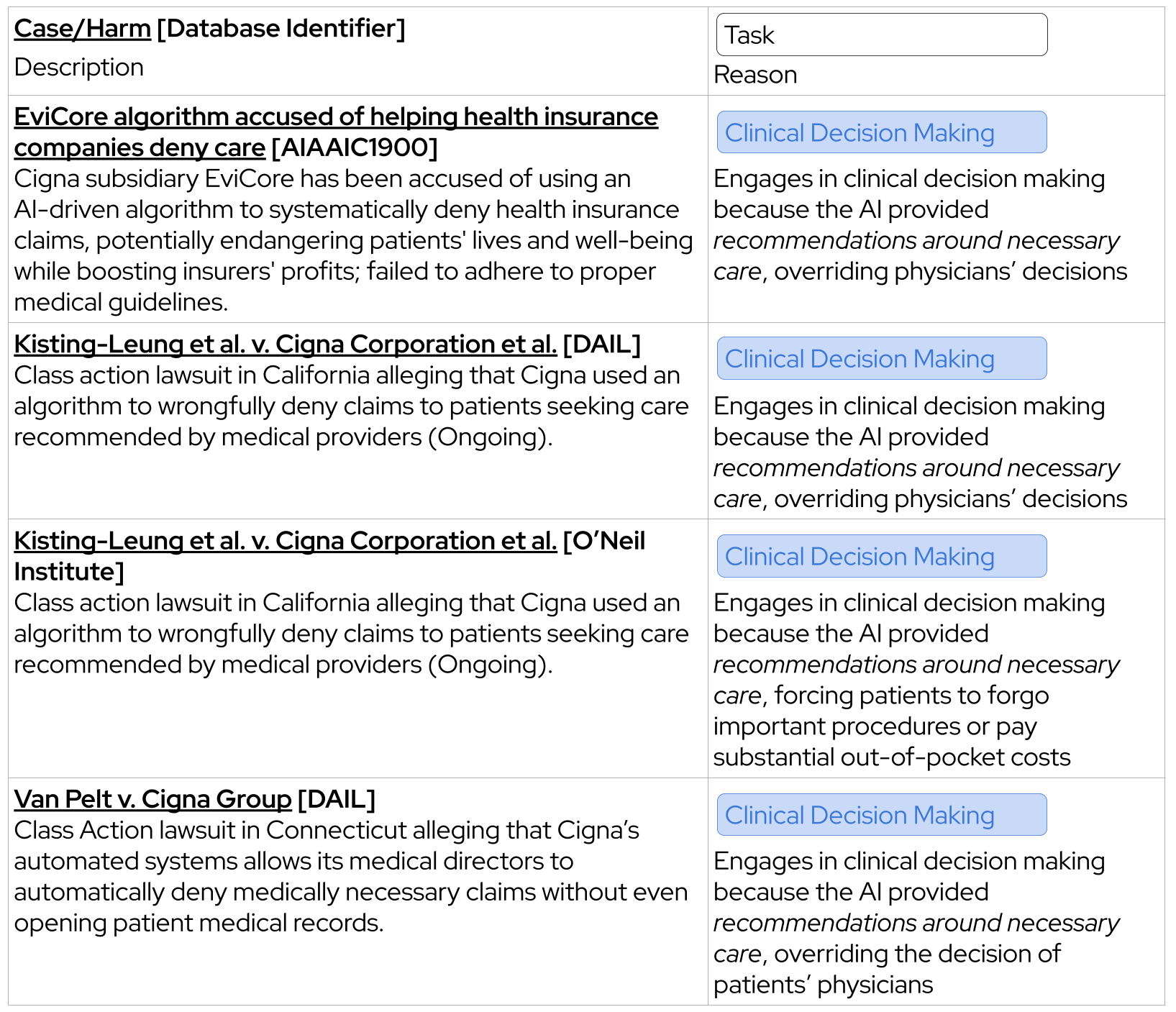} 
  \includegraphics[width=0.47\textwidth]{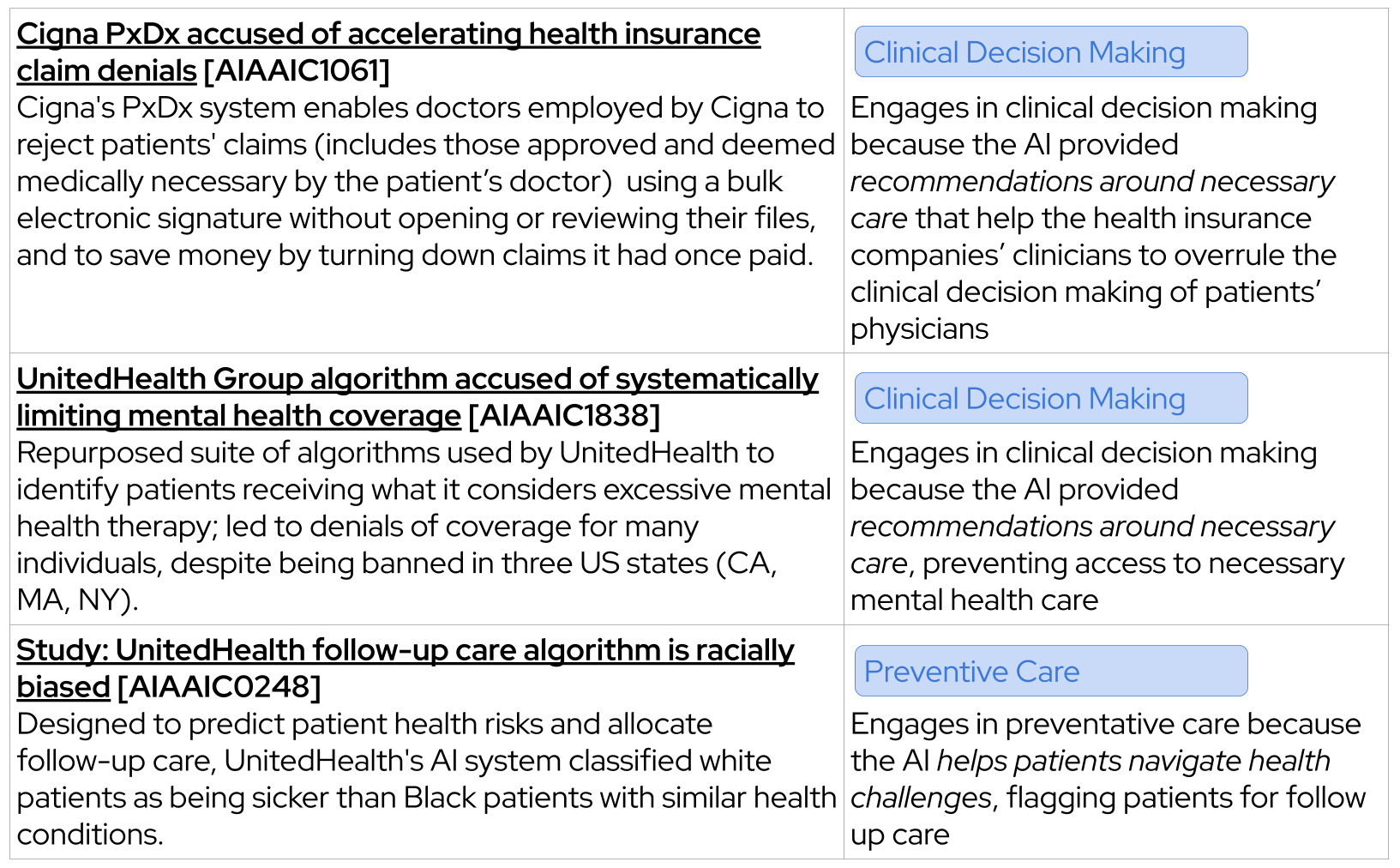} 
  \caption*{}
 \label{fig:LegalCases-Harms_3}
\end{figure}


\end{document}